\documentclass[times]{ettauth}

\usepackage{moreverb}

 \usepackage{amssymb}
  \usepackage{latexsym}
  \usepackage{amsfonts}
  \usepackage{amsthm}
  \usepackage{amsmath}
  \usepackage{graphics}
  \usepackage{setspace}
  \usepackage{graphicx}
  \usepackage{epstopdf}
  \usepackage{color}
  \usepackage{float}
  \usepackage{wrapfig}
  \usepackage{color}
  \usepackage{rotating}
  \usepackage{array} 
  \usepackage{dblfloatfix}
  \usepackage[hyphens]{url}
  \usepackage[table]{xcolor}
  \usepackage{multirow}

  \usepackage{graphics}
  \usepackage{setspace}
  \usepackage{algorithm}
  \usepackage{algorithmic}
  \usepackage{graphicx}
  \usepackage{epstopdf}  
  \usepackage{color}
  \usepackage{float}
  \usepackage{wrapfig}
  \usepackage{color}

\usepackage[colorlinks,bookmarksopen,bookmarksnumbered,citecolor=red,urlcolor=red]{hyperref}

\newenvironment{noindlist}
 {\begin{list}{\labelitemi}{\leftmargin=1.2em \itemindent=-.5em}}
 {\end{list}}

\newcommand\BibTeX{{\rmfamily B\kern-.05em \textsc{i\kern-.025em b}\kern-.08em
T\kern-.1667em\lower.7ex\hbox{E}\kern-.125emX}}

\def\volumeyear{2014}

\begin{document}

\runningheads{Perera et al.}{Sensing as a Service Model for Smart Cities Supported by Internet of Things}

\articletype{RESEARCH ARTICLE}

\title{Sensing as a Service Model for Smart Cities Supported by Internet of Things}

\author{A.~N.~Other\corrauth}

\author{Charith Perera$^{1,2}$,
 Arkady Zaslavsky$^{2}$,
Peter Christen$^{1}$,
Dimitrios Georgakopoulos$^{2}$}


\address{$^{1}$Research School of Computer Science, The Australian National University, Canberra, ACT 0200, Australia\\
$^{2}$CSIRO ICT Center, Canberra, ACT 2601, Australia}


\corraddr{CSIRO ICT Center, Canberra, ACT 2601, Australia. E-mail: charith.perera@csiro.au}

\begin{abstract}
The world population is growing at a rapid pace. Towns and cities are accommodating half of the world's population thereby creating tremendous pressure on every aspect of urban living. Cities are known to have large concentration of resources and facilities. Such environments attract people from rural areas. However, unprecedented attraction has now become an overwhelming issue for city governance and politics. The enormous pressure towards efficient city management has triggered various \textit{Smart City} initiatives by both government and private sector businesses to invest in ICT to find sustainable solutions to the growing issues. The Internet of Things (IoT) has also gained significant attention over the past decade. IoT envisions to connect billions of sensors to the Internet and expects to use them for efficient and effective resource management in Smart Cities. Today infrastructure, platforms, and software applications are offered as services using cloud technologies. In this paper, we explore the concept of sensing as a service and how it fits with the Internet of Things.  Our objective is to investigate the concept of sensing as a service model in technological, economical, and social perspectives and identify the major open challenges and issues.
\end{abstract}



\maketitle


\section{Introduction}
\label{sec:Introduction}

\begin{figure}[t]
 \centering
 \includegraphics[scale=.38]{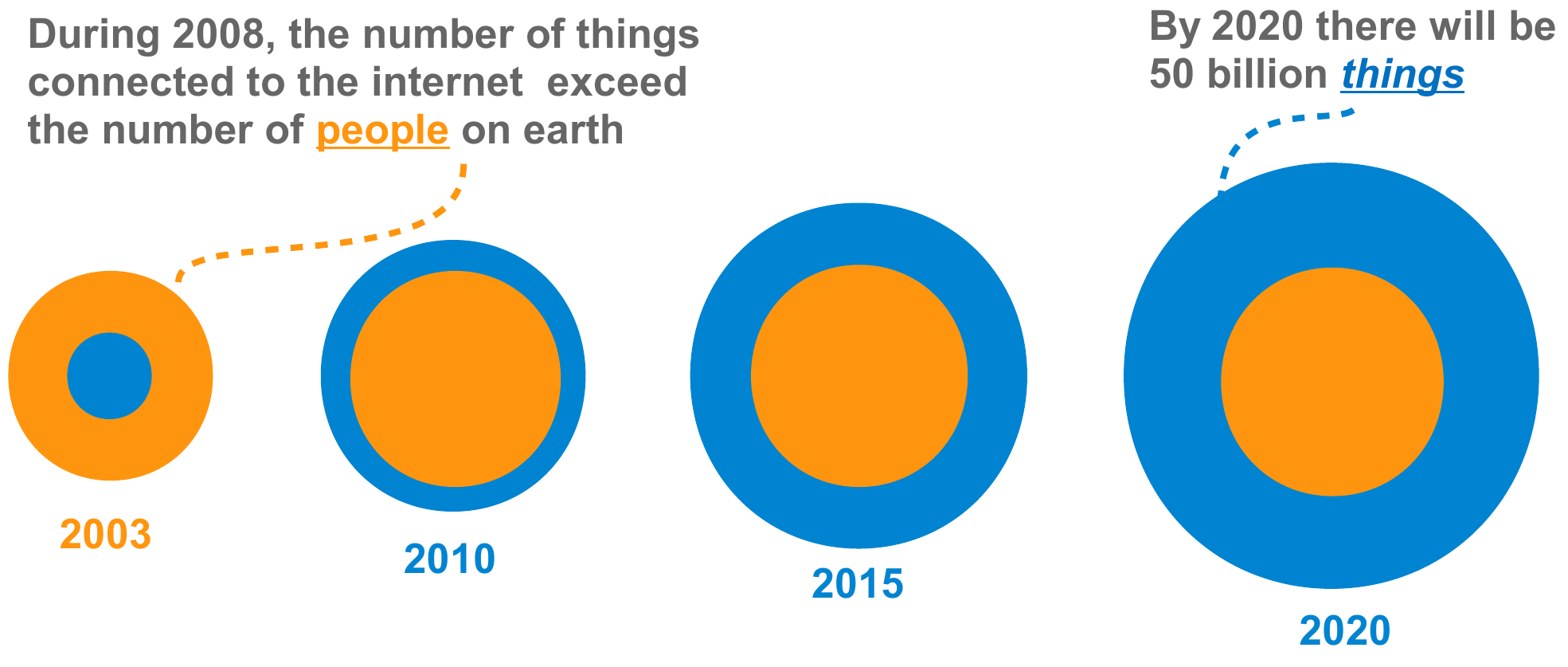}
 \caption{Growth of \textit{`things'} connected to the Internet \cite{P574}}
 \label{Figure:IoT_Statistics}	
\end{figure}

The Internet of Things (IoT) \cite{P001} and Smart Cities (SC) \cite{P532} are recent phenomena that have attracted the attention from both academia and industry. While both ideas consolidate similar ideology, they have different origins. Both IoT and SC do not have clear and concise definitions due to their short history and broadness. Examining the origins of both ideas in brief allows us to understand their potentials. Even though the term `\textit{Internet of Things}' was coined in 1999 \cite{P065}, the technologies that enable IoT such as sensor networks existed since the 1990s. Due to the advances in sensor and cloud technology, processing and storage capability, and decreased sensor production cost, the growth of sensor deployments has increased over the last five years \cite{ZMP003}. The European Commission has predicted that by 2020, there will be 50 to 100 billion devices connected to the Internet \cite{P029}. According to Figure \ref{Figure:IoT_Statistics}, the number of things connected to the Internet exceeded the number of people on earth in 2008.  

\begin{figure*}[t]
 \centering
 \includegraphics[scale=.80]{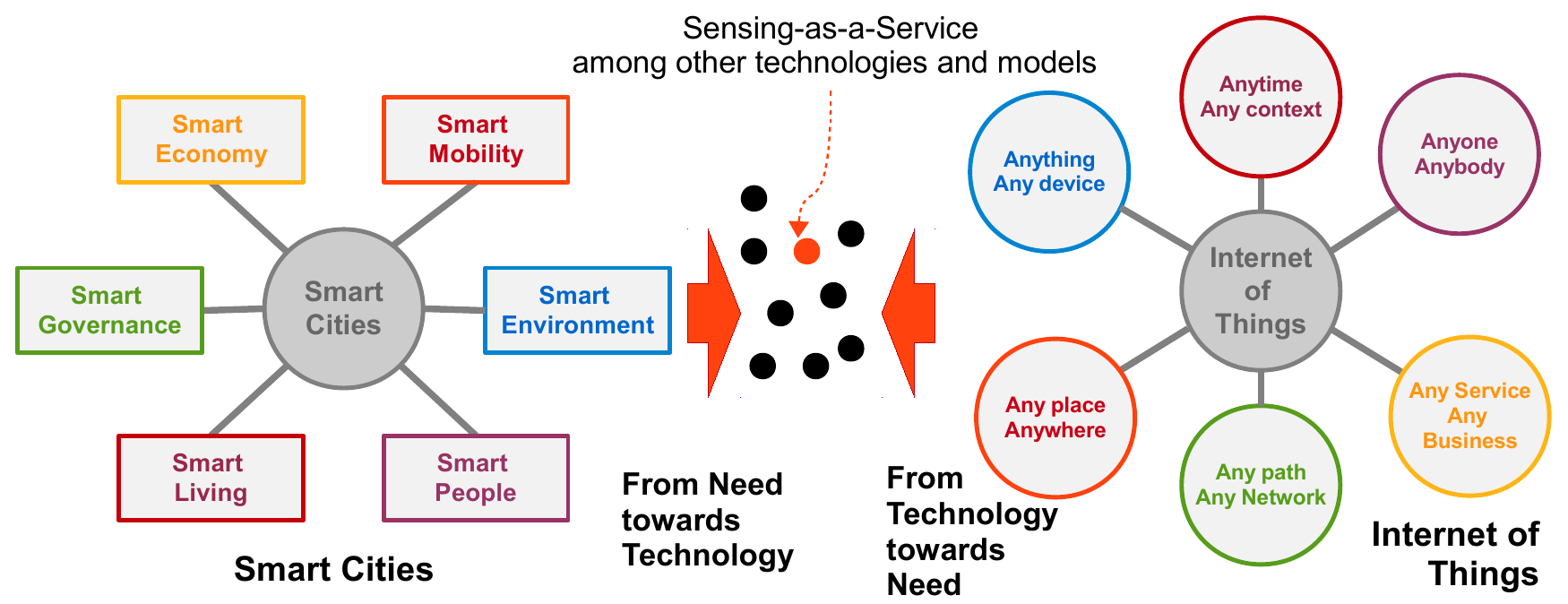}
 \caption{Relationship among sensing as a service model, SC and IoT}
 \label{Figure:Smart_Cities_and_IoT_Models}	
\vspace{-0.20cm}	
\end{figure*}

By definition, \textit{IoT allows people and things to be connected anytime, anyplace, with anything and anyone, ideally using any path/network and any service} \cite{P019}. As we can observe, IoT is primarily driven by technological advances, not by the applications or user needs. In contrast, SC \cite{P534} originated to solve the problems in modern cities. As a result of rural migration and suburban concentration towards cities, the urban living has become a significant challenge to both citizens and to the city governance. Waste, traffic, energy, water, education, unemployment, health, and crime management are some of the critical issues \cite{P535}. SC are expected to address these challenges efficiently and effectively using information and communication technologies (ICT). By definition, \textit{Smart Cities have six characteristics: smart economy, smart people, smart governance, smart mobility, smart environment and smart living} \cite{P528}. As illustrated in Figure \ref{Figure:Smart_Cities_and_IoT_Models}, SC and IoT, which have different origins, are moving towards each other to achieve a common goal. We believe that the sensing as a service model resides in between these two with many other technological and business models.

The remainder of this paper is organized as follows: we briefly review the trend of everything as service in Section \ref{sec:The_Trends}. In Section \ref{sec:Sensing_as_a_Service Model}, the sensing as a service model is presented. Subsequently, we explain the sensing as a service model using a futuristic scenario in Section \ref{sec:The Future}. In Section \ref{sec:Action}, we discuss several use case scenarios that highlight the different aspects of the sensing as a service model. Advantages in sensing as a service model are discussed in Section \ref{sec:Advantages}. Later, in Section \ref{sec:Challenges}, we highlight some of the major open challenges and issues related to sensing as a service model. Open challenges are identified under three main categories: technological, economical, and social. Finally, we present the concluding remarks in Section \ref{sec:Conclusions}.

\section{The Trends: Everything as a Service}
\label{sec:The_Trends}

Everything as a service (XaaS) \cite{P533} is a category of models introduced with cloud computing \cite{P498}. Similar to IoT, cloud computing also has a short history. It became popular with a number of industry initiatives such as Salesforce.com (1999) and Amazon Web Service (2002). The basic idea behind cloud computing is to concentrate resources such as hardware and software into few physical locations and offer those resources as services to a large number of consumers who are located in many different geographical locations around the globe over the Internet in an efficient manner. There are three major service models that are closely bound to cloud computing from its initial stage: infrastructure-as-a-service (IaaS), platform-as-a-service (PaaS), and software-as-a-service (SaaS). The commonality among these models is that they all provide resources as a service. With the popularity of these models, several similar type models are also proposed. The service models offered in cloud computing are discussed in \cite{P502} with popular industry based examples.

Let us briefly discuss the reasons behind the success of everything as a service model in the cloud paradigm. One major reason is the cost effectiveness. XaaS model promotes the \textit{`pay as you go'} method or in other terms \textit{`pay only for what you use'}.  This allows the consumers to consume a service from a service provider by paying only for the amount of resources they use. This is an efficient way compared to the traditional methods of consuming resources where consumers need to buy resources in predefined discreet quantities with higher expenses. For example, consider a retail online business which has peak and off-peak seasons. In traditional method, the business has to buy significant amount of compute servers (and other resources) to facilitate the customer needs during the peak season. However, these resources become idle during the off-peak season which makes the business process inefficient. In XaaS, online retail applications are hosted in servers facilitated by cloud service provider where the business is only required to pay for the resource it consumes. This model works similar to the utility services such as electricity. Further, cloud computing service models provide many other benefits such as business agility, scalability and elasticity, reliability, green initiatives, less maintenance work including backup and disaster recovery. Ultimately, XaaS allows businesses to focus more on core competency and innovation instead of ICT \cite{P539}. Further explanation on characteristics, features and benefits of cloud computing are presented in \cite{P498,P501}.

Smart City initiatives have become another trend during the past decade. Various city councils, business organizations, research and academic institutions, and the governments have invested significantly in projects to study, design, and build solutions to address the problems in urban cities using ICT. \textit{IBM Smart Planet and Smart Cities}, \textit{Oracle iGovernment},\textit{ Amsterdam Smart City}, \textit{Dubai SmartCity}, \textit{EuropeanSmartCities}, and\textit{ Smart Cities Future} are some of the leading Smart City projects \cite{P537, P536}. The following statistics show the magnitude of both trends. Global cloud computing and XaaS market is expected to grow from \$37.8 billion in 2010 to \$121.1 billion by 2015, growing at a compound annual growth rate (CAGR) of 26.2\% from 2010 to 2015 \cite{P539}. Similarly, the global Smart City market is expected to exceed \$1 trillion by 2016, growing at a CAGR of 14.2\% \cite{P358}.

\section{Sensing as a Service Model}
\label{sec:Sensing_as_a_Service Model}

\begin{figure*}[b]
 \centering
 \includegraphics[scale=.40]{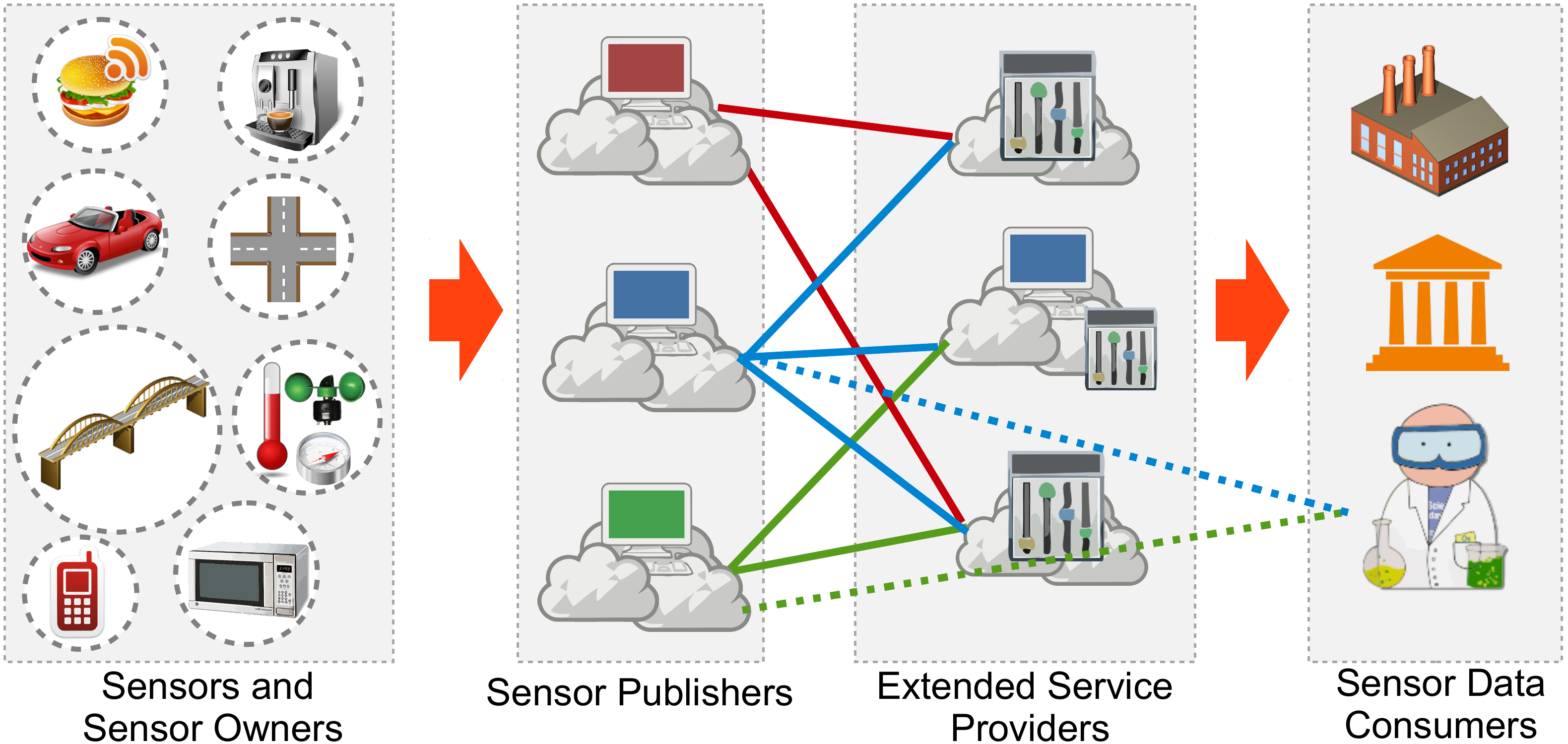}
 \caption{The sensing as a service model}
 \label{Figure:SENaas_Model}	
\end{figure*}

Previously, we introduced the sensing as a service model as a solution  based on IoT infrastructure. It has the capability to address the challenges in Smart Cities. As a result of getting 50 billion \textit{things} connected to the Internet by 2020, there will be many sensors available that can be used. Even today, many everyday objects are embedded with sensors though the usage is restricted to the object itself. Let us discuss the sensing as a service model and architecture in detail. As depicted in Figure \ref{Figure:SENaas_Model}, the sensing as a service model consists of four conceptual layers: 1) \textit{sensors and sensor owners}, 2) \textit{sensor publishers}, 3) \textit{extended service providers}, and 4) \textit{sensor data consumers}. In this section, we explain the sensing as a service model in a generic conceptual form. In Section \ref{sec:The Future}, we present a real world scenario based on this model. At the end of Section \ref{sec:The Future}, we map the real world scenario into the conceptual model in order to provide a practical understanding.

\begin{figure}[t]
 \centering
 \includegraphics[scale=.32]{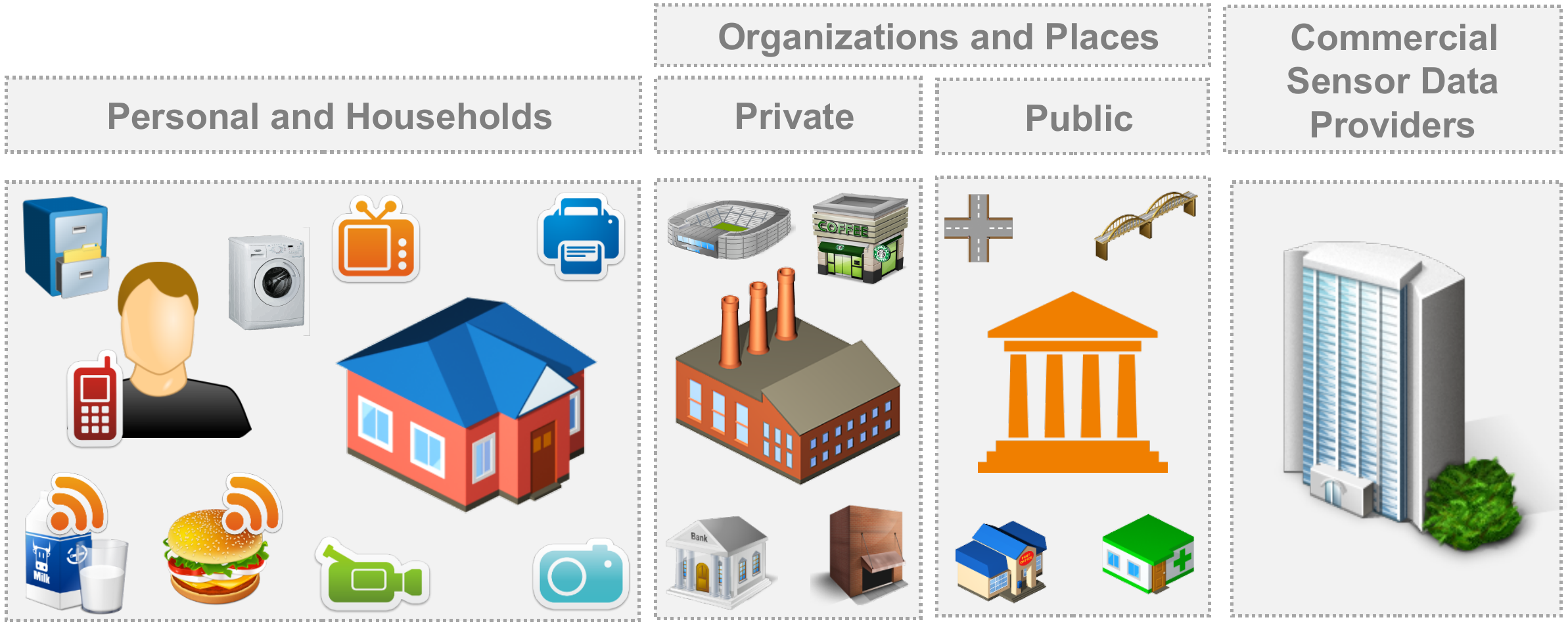}
 \caption{Sensor classification scheme based on ownership}
 \label{Figure:Classification_based_on_Sensor_Ownership}	
\vspace{-0.12cm}	
\end{figure}

\textbf{Sensors and Sensor Owners Layer:} This layer consists of sensors and sensor owners. A sensor is a device that detects, measures or sense a physical phenomenon such as humidity, temperature, etc. \cite{P009}. Multiple sensors can be attached to an object or device. For example, microwaves or coffee machines may have sensors that can be used to detect events (e.g. the number of times it is used per day and related context information). Such information can be used to understand user behaviour and user preferences more accurately. A road may have sensors that can detect the weather and traffic conditions. Today, large varieties of different sensors are available. They are capable of measuring a broad range of phenomena \cite{P595}. Further, they have the capability to send  sensor data to the cloud. On the other hand, a sensor owner has the ownership of a specific sensor at a given time. Ownership may change over time. We classify sensors into four categories based on ownership as depicted in Figure \ref{Figure:Classification_based_on_Sensor_Ownership}: \textit{personal and household}, \textit{private organizations / places}, \textit{public organizations / places}, and \textit{commercial sensor data providers}. In  addition to sensor data, related context information also has a significant value \cite{ZMP007}.

\begin{noindlist}
\item All  personal items, such as mobile phones, wrist watches, spectacles, laptops, soft drinks, food items and household items, such as televisions, cameras, microwaves, washing machines belong to the personal and household category. In simple terms, all  items (and also all sensors) not own by private or public organizations belong to this category. We expect that all of these items (also called things, objects, and devices) would be equipped with sensors in the future.

\item The private organizations and places category consists of all  items own by private organizations. The same items we listed under personal and household category can be listed under here as well depending on the ownership. If a private company owns a coffee machine and a microwave which cannot be attributed to a single person, then those items can be listed under this category. Therefore, the private business organization has the right to take the decision whether to publish the sensors attached to those items to the cloud or not. As another example, if a private business organization owns a sport complex or a hospital, all the sensors deployed in those properties are also owned by them. When a company manufactures and sells a product that comprises sensors, the ownership get transferred to that customer. As a result, a customer will decide whether to publish those sensors in the cloud or not. The same process will occur when physical properties (e.g. land, building) are sold from one party to another. This category would be the second largest sensor  owner after the personal and household category.

\item The public organizations and places category is similar to the private organizations and places category we discussed above. However, this category also includes public infrastructure such as bridges, roads, parks, etc. All the sensors deployed by the government will be published in the cloud depending on government policies.

\item Commercial sensor data providers are business entities who deploy and manage sensors by themselves by keeping ownership. They earn by publishing the sensors and sensor data they own through sensor publishers. They may deploy sensors across all places such as households, private and public owned properties depending on demand and strategic value by also complying with legal terms. Mostly, they will focus on public and private places. They will also make a payment to the property owner as an exchange for giving permissions for sensor deployment. For example, commercial sensor data provider may deploy sensors in a children's park owned by state government (under government permission) to detect motion and measure the micro climate (e.g. temperature, humidity, wind speed, wind direction). Such monitoring allows to detect and predict potential crowd movements. The sensor data that can be used to predict such movements can be sold to sensor data consumers such as mobile stall businesses and children's product retailers who may be located in nearby areas.

\end{noindlist}

A sensor owner makes the final decision on whether to publish the sensors he owns in the cloud or not. If the owner decides not to publish, no sensor publisher would be able to get access to those sensors which significantly protect the security and privacy of the sensor owner. If the sensor owner decides to publish the sensors he owns, he needs to register himself with a sensor publisher. Sensor owners can define restrictions and conditions such as who can request permission and the expected return (offer). It is important to note that each sensor may send data to a different SP in the cloud (similar as we use Internet service providers). However, a single sensor only sends data to a single SP (in order to save energy). Data will be shared between SPs if necessary depending on consumer requirements. Even though all four categories perfom the same task (i.e. sensor deployment and publication), the decision making processes can be quite different especially in term of objectives, financial goals, approval processes, privacy and policy concerns.

\begin{figure*}[t]
 \centering
 \includegraphics[scale=.65]{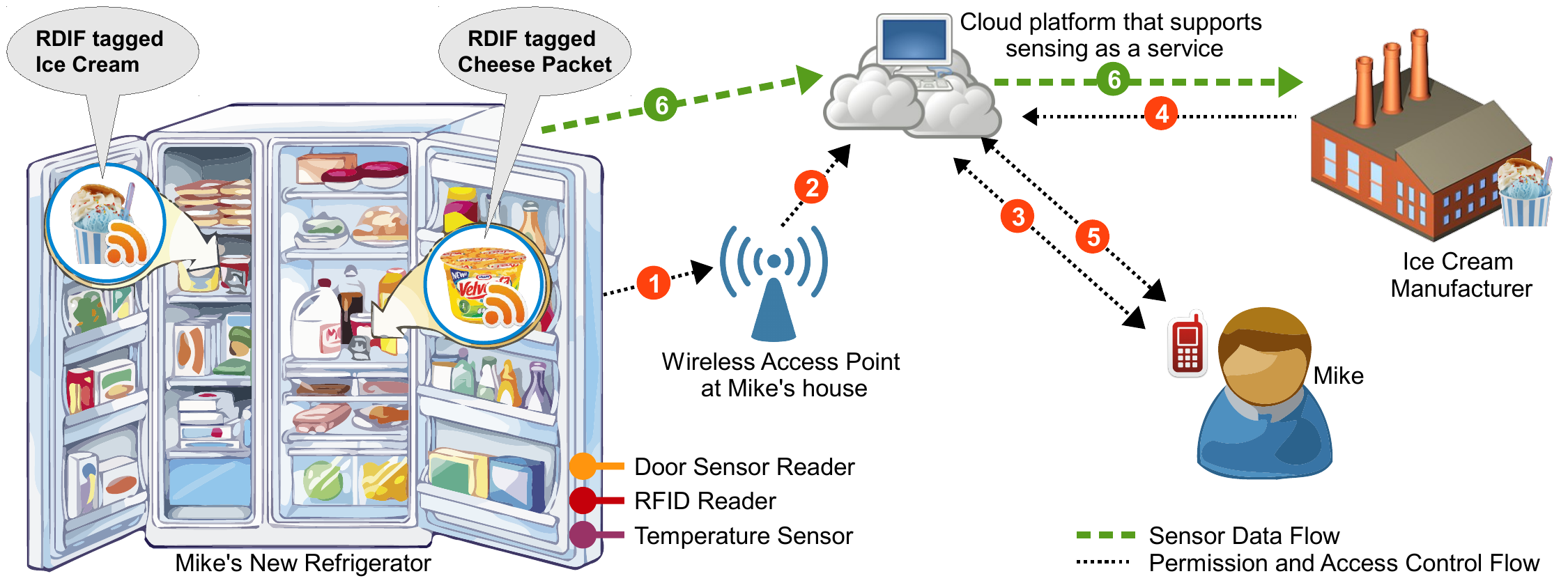}
 \caption{A futuristic scenario that explains the interactions in sensing as a service model}
 \label{Figure:UseCase}	
\vspace{-0.18cm}	
\end{figure*}

\textbf{Sensor Publishers Layer:} This layer consists of sensor publishers (SP). The main responsibility of a sensor publisher is to detect available sensors, communicate with the sensor owners, and get  permission to publish the sensors in the cloud. Sensor publishers are separate business entities. When a sensor owner registers a specific sensor, SP collects information about the sensor availability, owner preferences and restriction, expected return, etc. All this information needs to be published in the cloud. Once the registration is done, a SP waits until a sensor consumer makes a request. When a SP receives such a request, it forwards all the details including the offer to the corresponding sensor owner(s) to accept or reject. If the sensor owner accepts the offer, the corresponding sensor data consumer will be able to acquire  data from that sensor through the SP during the period mentioned in the agreement (offer). The same interaction explained above can take place between SPs and ESPs. SPs entirely depend on the payments (e.g. commission) receives from sensor owners, sensor data consumers or both. \textit{Xively} \cite{P579} is a public cloud for the IoT that simplifies and accelerates the creation, deployment, and management, of sensors in scalable manner. Further, it allows sharing sensor data with each other though it is far away from being qualified as a SP  we envisioned in the sensing as a service model. The \textit{OpenIoT} project \cite{P377}  focuses on providing an open source middleware framework enabling the dynamic formulation of self-managed cloud environments for IoT applications. Global Sensor Networks (GSN) \cite{P227} is a middleware which supports sensor deployments and offers a flexible, zero-programming deployment and integration infrastructure for IoT. These approaches strengthen our vision towards sensing as a service.

\textbf{Extended Service Providers Layer:} This layer consists of extended service providers (ESP). This layer can be considered as the most intelligent among all the four layers which embed the intelligence to the entire service model. The services provided by ESPs can be varied widely from one provider to another. However, there are some fundamental characteristics of ESPs. To become an ESP, they have to provide value added services \cite{P668} to the sensor data consumers. However, in some instances a single business entity can perform both sensor publisher and extended service provider roles. Each SP has access (only) to the sensors which are registered with it. When a sensor data consumer needs sensor data from multiple sensors where each sensor has been registered with different SPs, ESPs can be used to acquire data easily. ESPs communicate with multiple SPs regarding sensor data acquisition on behalf of the sensor data consumer. The ESPs depend on the payments (e.g. commission) similar to SPs. ESPs receive payments for the value added service they provided to their customers (i.e. sensor data consumers). An example value added service can be selecting sensors based on customer's requirements \cite{ZMP004}. Customers will provide their requirements in high-level (e.g. measure environmental pollution in Canberra) instead of selecting the sensors by themselves. In return, ESP will select the appropriate sensors (e.g. pH, temperature, humidity, CO$_{2}$, etc.) located in Canberra. Pinto et al. \cite{P666} have proposed  an architectural approach for telecoms to take advantage of machine-to-machine markets in the IoT domain. It explains the opportunities  business can address by providing services related to connectivity management, data management, and service provisioning.

\textbf{Sensor Data Consumers Layer:} This layer consists of sensor data consumers. All the sensor data consumers need to register themselves and obtain a valid digital certificate from an authority in order to consume sensor data. Some of the major sensor data consumers would be governments, business organizations, academic institutions, and scientific research communities. Sensor data consumers do not directly communicate with sensors or sensor owners. All the communication and  transactions need to be done through either SPs or ESPs. If a sensor consumer has the required technical capability, they can directly acquire data from sensor publishers. However, this could be very challenging. For example, selecting which sensors to use out of billions of sensors available could be an overwhelming task \cite{ZMP006}. Further, sensor data consumers may need to communicate with multiple sensor publishers to acquire the required data. However, the cost of sensor data acquisition would be lower as they are not required to pay for ESPs' value added services. Scientific research communities may be interested in such methods. The sensor consumers with less technical capabilities and expertise can acquire required sensor data through ESPs where most of the difficult tasks such as combining sensor data from multiple sensor publishers and selecting appropriate sensors based on the consumer requirements are handled. Further, sensor consumers can register their interests with both SPs, and ESPs. For example, they can express their interest by using a number of constraints. A coffee manufacture who expects to starts its business in Canberra may be interested to access the sensor data produced by coffee machines located in Canberra for a fee. Depending on the expression of interest, ESPs/SPs will notify the coffee manufacturer when a matching deal is available. In simple terms, sensor owners define what they are expecting as return for the sensor data from one end of the Sensing as a service model. On the other end, sensor consumers define what kind of sensor data they want and how much are they willing to pay (offer). SPs and ESPs are platforms that enable these transactions (deals) to take place. The sensing as a service model shares common characteristics of an auction \cite{P671}.

\begin{figure*}[t]
 \centering
 \includegraphics[scale=.68]{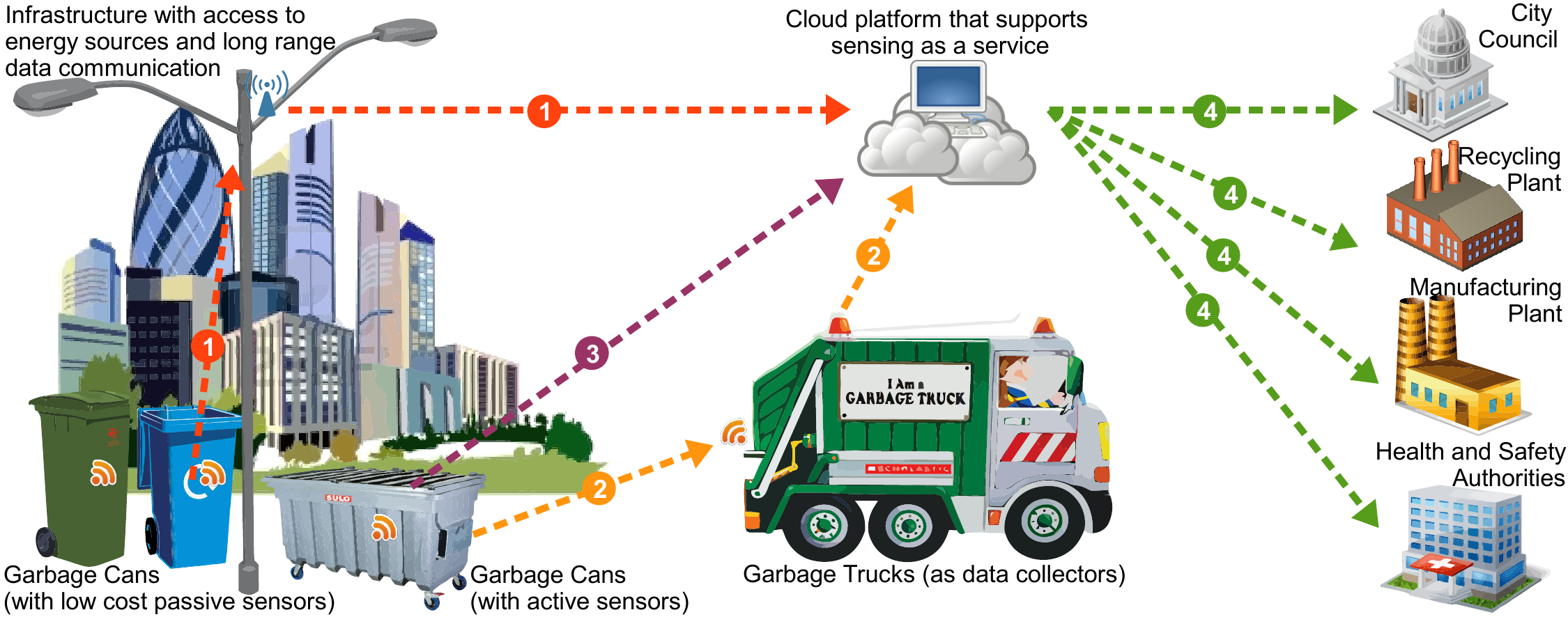}
 \caption{Efficient waste management in Smart Cities supported by the sensing as a service model}
 \label{Figure:UseCase3}	
\vspace{-6pt}
\end{figure*}

\vspace{-6pt}

\section{The Future: A Scenario}
\label{sec:The Future}

A futuristic scenario can be used to explain the sensing as a service model. The scenario illustrated in Figure \ref{Figure:UseCase} is based on smart home domain which also plays a significant role in the Smart Cities. Our intention is to highlight the interactions between different parties explained earlier in high-level.
 
\textit{Mike} bought a new refrigerator for his new home. He brought it home and plugged it to the power. The fridge automatically identifies the availability of Wi-Fi in the house as shown in step (1). Further, the refrigerator communicates with a sensor publisher and informs about its presence by providing information such as the available sensors (e.g. RFID reader, temperature, door sensors) as shown in  step (2). 
Next, in step (3), the SP communicates with Mike to check whether he likes to publish the sensors attached to the refrigerator in the cloud (step 3). We assume that \textit{Mike} has already registered with the SP in a previous transaction. \textit{Mike} is allowed to define which sensors to publish, what kind of consumers are allowed to bid, and what kind of return (fee or any other offer) is expected. Later, \textit{Mike} receives an email from a company called \textit{DairyIceCream} (via a SP called \textit{EasySensing}), an ice cream manufacturer, with an offer as shown in  step (4). \textit{DairyIceCream} is interested to have access to the RFID reader and the door sensor attached to the freezer in  \textit{Mike's} refrigerator. As a return, \textit{DairyIceCream} is willing to offer either 3\% discount on every product purchased from \textit{DairyIceCream} or a monthly fee of \$2. As \textit{Mike} likes \textit{DairyIceCream} products, he agrees to the 3\% discount offer instead of the monthly fee as shown in  step (5). A week later, \textit{Mike} receives an email from a company called \textit{ProductiveAnalytics} which has been sent on behalf of the \textit{GoldenCheese} company, a cheese manufacturer, with an similar offer. This request  also comes through \textit{EasySensing}. However, the offer is either 4\% discount on every product purchase by \textit{GoldenCheese} or a monthly fee of \$1. As \textit{Mike} does not like \textit{GoldenCheese} products, he decides to accept the monthly fee option. 


\textbf{Scenario from model perspective:} In Section \ref{sec:Sensing_as_a_Service Model}, we explained the sensing as a service model in a generic perspective and now we describe it from the above mentioned scenario perspective. In the scenario, \textit{Mike} is the sensor owner. Therefore, he and his sensors represent the \textit{sensors and sensor owners} layer. Further, in ownership categorization, \textit{Mike} represents the \textit{Personal and households} scheme. Both the \textit{DairyIceCream} and \textit{GoldenCheese} companies represent the \textit{sensor data consumers} layer. \textit{EasySensing} is a SP who enables the communication and transactions between \textit{Mike} and the \textit{DairyIceCream}. \textit{EasySensing} is responsible for matching the sensor owners expectations  with the requirements of sensor data consumers. \textit{DairyIceCream} retrieves the data from \textit{EasySensing} directly and conducts the data analysis with the help of in-house experts. \textit{ProductiveAnalytics} is an ESP who works on behalf of \textit{GoldenCheese}. \textit{GoldenCheese} has hired \textit{ProductiveAnalytics} to perform the data analysis as they do not have the required technical skills within the company. \textit{ProductiveAnalytics} collects the data by handling all the deals and transaction with the sensor owners though their partner SPs.


\section{Sensing as a Service in Action}
\label{sec:Action}

In the previous section, we discussed a scenario related to the smart home domain in sensing as a service perspective. This section presents three different use case scenarios that explain different aspects of the sensing as a service model: (1) waste management, (2) smart agriculture, and (3) environmental management. All three scenarios share common a sets of characteristics as well as few unique characteristics. Waste management has a direct impact on cities. Environmental management has  direct, indirect, and long term impact on the entire human life-cycle both in urban and rural living. Further, smart agriculture makes indirect impact on sustainability towards the smart cities.

\subsection{Waste Management}
\label{sec:Action:Waste_Management}

Waste management is one of the toughest challenge that modern cities have to deal with. Waste management consists of different processes such as  collection, transport, processing, disposal, managing, and monitoring of waste materials. These processes cost significant amount of money, time, and labour. Optimizing waste management processes help to save money that can be used to address other challenges that smart cities need to deal with. In Figure \ref{Figure:UseCase3}, we illustrate how the sensing as a service model works in the waste management domain. In a modern smart city, there are several parties who are interested in waste management (e.g. city council, recycling companies, manufacturing plants, and authorities related to health and safety). Instead of deploying sensors and collecting information independently, the sensing as a service model allows all the interest groups to share the infrastructure and bare the related costs collectively. The most important aspect of such a collaboration is the cost reduction that individual groups need to spend otherwise. All the interested parties can retrieve and process sensor data in real time in order to achieve their own objective. The cost depends on the data requirement of the interest group.

For example, a city council may use sensor data to develop optimized garbage collection strategies, so they can save fuel cost related to garbage trucks. Additionally, recycling companies can use sensor data to predict and track the amount of  waste coming into their plants to be processed so they can optimize their internal processes. Further, health and safety authorities can monitor and supervise the waste management process without spending substantial amount of money for manual monitoring inspections. The phenomenon of sharing sensor data using a sensing as a service model creates a synergy effect (i.e. interaction of multiple elements in a system to produce an effect greater than the sum of their individual effects). The sensing as a service model ensures the long term sustainability of the IoT infrastructure.

Let us discuss how this technology can be used to support the sensing as a service model in financially viable manner. In order to perform waste management, different types of sensors need to be deployed in different places such as garbage cans and trucks. These sensors need to detect information such as the amount of garbage, types of  garbage, and so on. As we have depicted in Figure \ref{Figure:UseCase3}, direct and indirect communication strategies can be used to collect and communicate sensor data to the cloud. Sensors with energy harvesting capabilities are important in this domain \cite{P633}. As represented in step (1) in Figure \ref{Figure:UseCase3}, low powered \cite{P634} and low capable sensors can be used to sense and data can be uploaded to the cloud with the help of nearby infrastructure (e.g. through communication devices attached to street lights or similar infrastructure that have access to rich energy sources and communication capabilities). Additionally, when long range communication is not available, data can be uploaded to the cloud with the help of auto-mobiles, as depicted in step (2) in Figure \ref{Figure:UseCase3}, such as garbage trucks, city council vehicles, buses that operate in the areas and so on. Furthermore, both active and passive sensors can be used to sense the environment \cite{P017}. Direct communication can be done via technologies such as 3G  which makes this approach less dependant on third parties (as depicted in (3) in Figure \ref{Figure:UseCase3}).

\subsection{Smart Agriculture}
\label{sec:Action:Smart_Agriculture}

\begin{figure}[t]
 \centering
 \includegraphics[scale=.61]{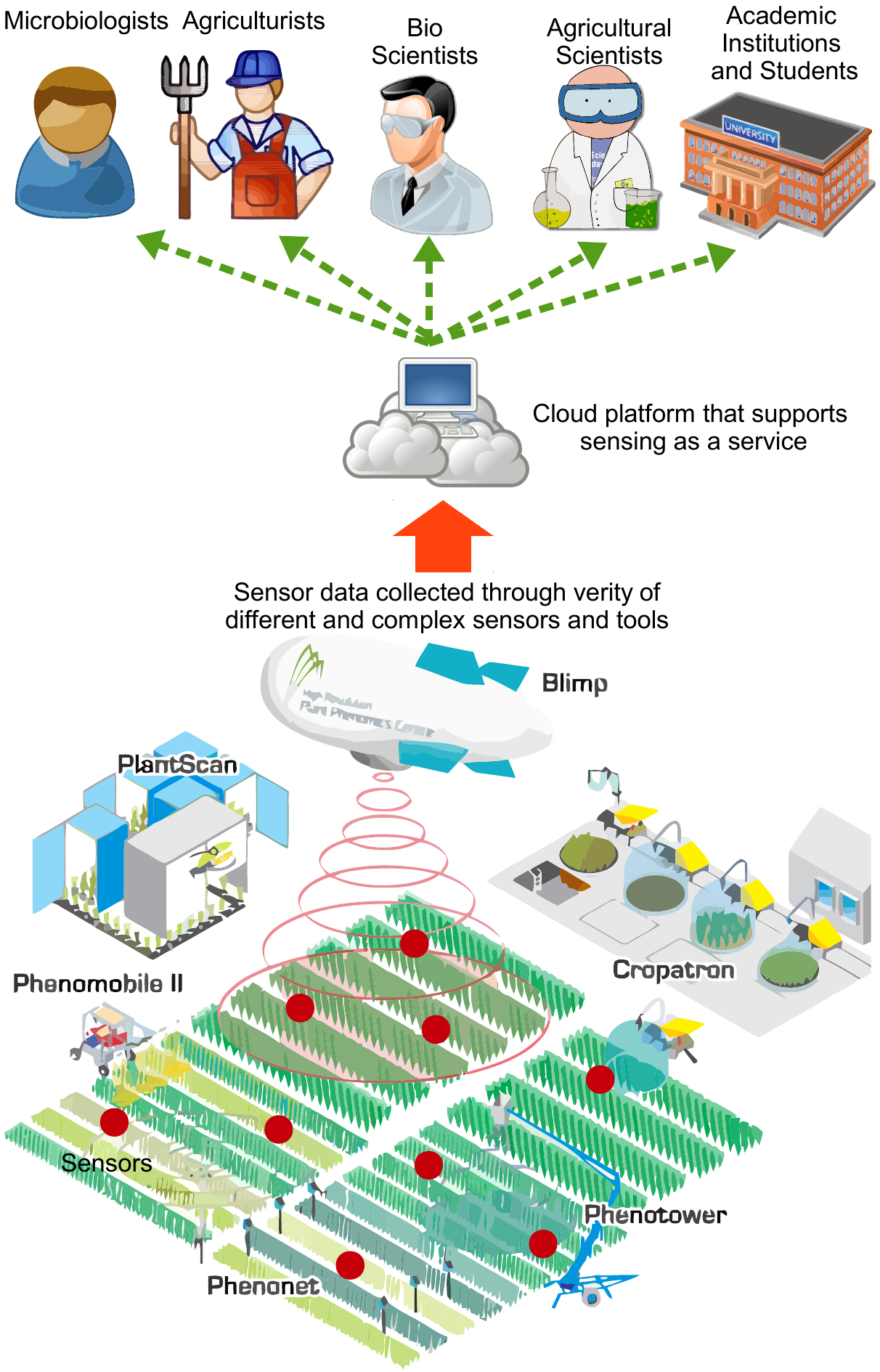}
 \caption{Efficient and effective collaborative research supported by sensing as a service model}
 \label{Figure:UseCase2}	
\end{figure}

Currently, the authors are actively involved in designing and developing open platforms for sensor data collection, processing and sharing in the domain of agriculture through two projects: \textit{Phenonet} \cite{P412} and \textit{OpenIoT} \cite{P377}. In this scenario, the general public is not directly involved as in the smart home domain. In Figure \ref{Figure:UseCase2}, we illustrate how the sensing as a service model works in the smart agriculture domain. Agriculture is an importation part of  smart cities as it contributes to the food supply-chain that facilitates a large number of communities concentrated into cities.

The sensing as a service model allows to conduct scientific research and exploration more efficiently and efficiently. Further, it opens up different research opportunities which are unlikely to occurr in a traditional research model. Let us explain the \textit{Phenonet} project in details and the applicability of the sensing as a service model towards agricultural research. \textit{Phenonet} describes the network of sensors collecting information over a field of experimental crops. Researchers at the High Resolution Plant Phenomics Centre are testing a network of smart sensor nodes able to monitor plant growth and performance information and climate conditions. Even though the main research goal of deploying sensors and collecting data is to understand plant growth under different climate conditions, the same set of sensors can be utilized to perform a verity of different research activities in different domains. The data can be shared among different research organizations and institutions located around the world. Due to limited funding, most of these research institutions may not be able to maintain large scale sensor deployments (e.g. academic institutions, specially in developing countries). However, the sensing as a service model allows all these interest groups, who are unable to set-up their own sensor deployments, to perform research using actual data with significantly less costs. Further, the sensing as a service model creates opportunities across different domains. For example, the above mentioned sensor data can be used to understand pest control and related phenomenon. Additionally they can be used to understand soil conditions where bio-scientist may be interested. More importantly, the  sensing as a service model allows researchers to share resources across  borders and understand phenomenon which are not available in their own countries.

\subsection{Environmental Management}
\label{sec:Action:Environment_Pollution_Management}

This domain has the unique ability of utilizing existing sensors that are deployed for different reasons. Most of the sensors used in environmental monitoring are commonly used in other domains such as climate, wild fire detection, and structural health monitoring. Using the sensing as a service model, interest groups can acquire relevant sensor data without deploying sensors by themselves. Further, environmental management is a large domain where a single organization cannot deal with (e.g. wild fire). A model like sensing as a service stimulates innovative solutions that use the same data but produce different results using different processing and analysing techniques (e.g. prediction, visualization, simulation). As we discussed in Section \ref{sec:Sensing_as_a_Service Model}, ESPs can help the sensor data consumers to orchestrate existing services into different data processing \cite{P672} and analysis work-flows \cite{P635}.

\section{Advantages and Benefits}
\label{sec:Advantages}

Some of the major advantages and benefits in the sensing as a service model are discussed below:

\begin{noindlist}
\item \textit{Built-in cloud computing}: It is modelled around cloud computing. Therefore, it inherits all the benefits of the fundamental cloud computing models such as IaaS, PaaS, and SaaS. Scalable and widely accessible processing and storage resources are available to facilitate sensing as a service software platforms (SPs and ESPs). Sensor data consumers only need to pay for the data they use. Therefore, the cost of data acquisition reduces significantly due to sharing, participatory / crowd sourcing,  and reusing nature (i.e. sense once and use by many). 

\item  \textit{Participatory sensing}: The workload is distributed among different players in the model. This enables rapid deployment of sensors across wider geographical locations that capture various phenomena.

\item  \textit{Sharing and reusing}: In traditional methods, each party (group or person) who wants to collect sensor data needs to visit the field and deploy the sensors manually by themselves. Further, there is no easy way to share  sensor data collected by one party with others. Sensing as a service is a model that stimulates by concept of sharing. In simple terms, if someone has already deployed the sensors, others can have access to them by paying a fee to the sensor owner. One of the major arguments that could arise regarding sensing as a service model is that \textit{``How to convince a manufacturer to embed sensors and communication capabilities into devices we use in everyday life (e.g refrigerator in the use-case presented in Section \ref{sec:The Future})''}. This question can be answered in two different perspectives. 

First, IoT envisions to have sensor embedded into objects around us. The goal of IoT is to allow devices to communicate with each other. Naturally, such a goal forces next generation devices to be embedded with rich sensing and communication capabilities. Therefore, the motivation is given to the manufacturers not by the sensing as a service model but the vision of IoT.  The sensing as a service model is designed to provide incentives to users which motivate them to purchase next generation devices that supports both IoT envisioned interactions as well as the sensing as a service model. The additional cost that contributes to increase the prices of the devices (due to embedding rich sensing and communication capabilities) can be easily covered by participating in the sensing as a service model itself. Even today, state of the art devices such as refrigerators and televisions comprise communication and sensing capabilities.

\item  \textit{Reduction of data acquisition cost}: Due to the shared and collaborative nature,  data acquisition cost will be reduced significantly. Such a sustainable business model stimulates more and more sensor deployments. Further, technological advances and higher demands allow to produce sensors in mass volumes using cheap materials by reducing the cost per unit. Further, this helps to  collect  data from sensors which was impossible previously.

\item  \textit{Collect data previously unavailable}: This model allows to collect sensor data which is impossible to collect using traditional non-collaborative methods. This business model promotes and stimulates the sensor deployments by companies at commercial level. As we explained earlier in Section \ref{sec:Sensing_as_a_Service Model}, dedicated business entities will deploy sensors in public places such as parks and bridges so  government authorities can have access to those sensors by paying only for the data they need in real-time or archived. Today business entities spend substantial amount of money to conduct market analyses and consumer surveys. A sample of 1,000 respondents, which would give a statistical accuracy of +/-3.1\% costs around \$8,000 \cite{P632}. Recently, different third party companies started offering consumer surveys on behalf of businesses. One such solution is Google Consumer Surveys (www.google.com/insights/consumersurveys). Google Consumer Surveys allows businesses to target user groups with specific criteria and conduct the survey. Currently, one user response cost around \$0.10, 1/10th of the cost of similar quality research conduct using traditional methods. 

Even though such approaches have reduced the cost of surveys, they still have deficiencies such as latency, inaccuracies, and so on. In the sensing as a service model, all the data is directly coming from the sensor without user intervention. This also helps to reduce the cost of data acquisition. Due to privacy concerns it is important to anonymise the sensor data collected. We discuss privacy matters later. In the smart home scenario we discussed in Section \ref{sec:The Future}, we explained how a  single sensor attached to a refrigerator, and cheap passive RFID tags attached to consumer products, produce valuable information of consumer behaviour that can be used by thousand of companies. This drastically reduces the consumer survey cost as well as pay off the cost of attaching sensors to the products.

\item  \textit{Innovations}: Due to a reduction in sensor data acquisition cost, larger number of interest groups will be able to access to them. Further, the availability of sensor data which was not available previously can also significantly stimulate innovation . Sensing as a service model itself provides space for innovation in the ESP layer. The cloud-based value added services provided in the ESP layer allows the sensor data consumers to achieve their objective easily and faster in many different application domains.

\item  \textit{Applications}: Easily accessible sensor data allows government authorities, academia, research institutions, and businesses to address different challenges in Smart Cities such as traffic, energy, water, education, and unemployment, health, and crime management. For example, accurate data on energy consumption in a city allows managing electric grids efficiently by analysing and predicting energy consumption behaviours, patterns, future trends, and needs.

\item  \textit{Real-time data for decision making and policy making}: This model enables collecting sensor data in real-time, from a variety of different domains, which facilitates the decision making processes. Such data is expensive to collect and usually unavailable for decision making in traditional sensor deploying environments. For example, data collected from sensors deployed in vehicles and roads allow the authorities to monitor and manage traffic in real-time. Further, sensor data collected over a period of time (archived) can be used to make policy decisions. For example, traffic data over a period on a specific city will help a city governance to make long term strategic decisions such as whether to invest on a tram service across the city or not. In addition to the points discussed above, there are many other direct and indirect benefits in the sensing as a service model.

\item  \textit{Direct and indirect benefits}: The sensing as a service model creates a win-win situation for all the parties involved. Based on the scenario we presented in Section \ref{sec:The Future}, \textit{Mike} (sensor owners' perspective) is getting a return (a valuable offer). In \textit{DairyIceCream}  perspective, now they have real-time data about product consumer behaviour (e.g. when \textit{Mike} eats ice cream, how frequent, whether \textit{Mike} use substitutions and so on). Therefore, \textit{DairyIceCream} is no longer required to conduct manual surveys and market analyses.

\item  \textit{Privacy preservation}: Finally and more importantly, this model provide complete control of the privacy of sensor owners in  their own hands. The final decision of whether to publish their sensors or not is taken by the sensor owners. It allows the sensor owners to control and protect their privacy. Additionally, the sensing as a service model needs to be supported by anonymization techniques. For example, lets consider security and privacy challenges \cite{P636} related to the  smart home scenario we presented in Section \ref{sec:The Future}. During the configuration process, it is important to identify the information and preferences related to \textit{Mike}. In order to protect the  privacy of the users, SPs and ESPs should not provide personal information to the sensor data consumers. Such approach helps to preserve user privacy. Additionally, once the deal between the sensor owner, sensor consumer and the sensor provider is done, data retrieves from  \textit{Mike's} sensors should be explicitly anonymized. It is important to develop new algorithms and security devices that can anonymize sensitive information (such as exact location).


\end{noindlist}


\begin{table*}[t]
\centering
\small

\renewcommand{\arraystretch}{-4}
\caption{Open challenges and issues in sensing as a service model}
\vspace{-0.3cm} 
\begin{tabular}{ m{0.05cm} m{2.0cm} m{12cm}  }

\hline  
   \multicolumn{1}{r}{}   &    
\begin{center} Open challenges and issues \end{center} & 
\begin{center} Description, significance, and research directions to address the challenges
 \end{center}

\\ \hline \hline


\begin{sideways}Technological\end{sideways}     
&   \begin{noindlist}
 		 \item Architectural Designs, \newline Sensor Configuration, \newline Data Fusing / Filtering,
 		 \newline Processing / Storage, \newline
 		 Infrastructure, \newline Energy Consumption,\newline
 		 \vspace{60pt}
 		 \item Standardization, \newline  Accuracy \newline Security and Privacy,  

    \end{noindlist}  
 &  \begin{noindlist}
  		 \item Technology plays the most important role in enabling the sensing as a service model. This model uses the same infrastructure that IoT envisions. Therefore, most of the technological solutions that are developed to facilitate sensing as a service can be used to realize the vision of IoT. The sensing as a service model is expected to facilitate billions of sensors and parallel sensor data streams. A major challenge is to develop middleware solutions that allow to handle such demand \cite{P377}. Similarly, this model needs significant improvements in data communication bandwidth \cite{P667} over the existing infrastructure (e.g. fiber). Another major challenge is the sensor configuration. The term \textit{`sensor configuration'} encapsulates different aspects of configuration that needs to be done: sensor embedded software, intermediate devices, and  cloud (middleware) software. In reference to the scenario we presented in Section \ref{sec:The Future}, sensors in \textit{Mike's} new refrigerator need to be configured autonomously so they can communicate with the SP. Such an approach needs to deal with challenges such as heterogeneity: sensor types (e.g. RFID, temperature), protocols and communications technologies (e.g. Wi-Fi, Zigbee). In addition, once a deal is done, sensor behaviour need to be configured according to the agreement between the sensor consumer and the sensor owner (e.g. sampling rate, data communication frequency and so on). Further, SPs and ESPs may need to configure their cloud software accordingly. The sensing as a service model is a distributed system. It is critical to utilise computational devices with different capabilities and capacities (e.g. sensors, mobile phones, Raspberry Pi, laptops, servers) \cite{ZMP005}. Another challenge is to ensure the interoperability among different sensor hardware and cloud solutions. Complying with common standards in key areas in the architecture (such as communication interfaces and data formats) is critical. Energy conservation is also a challenge that needs to be addressed across all the entities in the model due to the large scale and the resource restricted nature of the sensors. Other than the sensor data, it is important to capture context information (e.g. battery level of the sensors, redundant sensors, access to energy sources, accuracy, reliability) as well  \cite{ZMP007}. Context information allows to design optimized sensing schedules and strategies that ensure the sustainability of the IoT infrastructure.

  		 \item Standardization is the key to interoperability. We have experienced the value of interoperability in service computing and many other occasions throughout the history of computing. Standardization efforts need to be carried out as early as possible to avoid significant frustrations and costs that may occur at later stages. Technology needs ensure the accuracy of the data up to an acceptable level as it is one of the main motivations behind the Sensing as a service model. It is important to anonymize the sensor data collected. Sensitive information such as location need to be implicitly altered to protect the sensor owner privacy. This should be done in both the hardware and software levels. For the hardware level, we need to develop next generation security appliance that can be used to anonymize data at the ground level (i.e. physically close to the sensor owners). Techniques similar to privacy preserving data sharing ad matching \cite{P637} need to be developed in order to combine sensor data to anonymize entities / profiles (excluding sensitive data) later at the server level. 
     \end{noindlist}  

\\ \hline
\begin{sideways} Economical  \end{sideways}      
&   \begin{noindlist}
		 \item Innovation, \newline Entrepreneurship, \newline Entry Barriers\newline
		 \vspace{30pt}
 		 \item Sustainability,\newline Licensing,\newline Business Practices, \newline Credibility
\end{noindlist}  
 &
  \begin{noindlist}
 		 \item The sensing as a service model will collect enormous amount of data that need to be processed and understood. It will  open up opportunities for thousands of new businesses. The entry barriers need to be kept at a minimum  to stimulate new start-ups to be established to provide more value added services (e.g. search sensors based on context information \cite{ZMP004} and user requirements \cite{ZMP006}). The opportunities are ranging from the point where data is collected and to the point data is delivered. As we argued earlier, most of the users who may consume sensor data will not have technical expertise. Therefore, understanding data and extract valuable information from sensor data, by data fusing and reasoning, can also provide value added services.
 		 
  		 \item The sensing as a service model promotes a healthy competition among parties involved as it helps both the sensor data owners and sensor data consumers. Sustainability needs to be ensured by having a fair and transparent financial model which motivates all the parties to be retained in the business. Sensor data and knowledge produced using them need to be accurate and credible so consumers can make important and potentially costly strategic decisions based on them.
 \end{noindlist}  

\\ \hline

\begin{sideways} Social  \end{sideways}      
&   \begin{noindlist}
		 \item Trust, \newline Social Acceptance, 
		 \newline Change Management, \newline Awareness 
		 		 
 		 \item Security and Privacy, \newline Safety, \newline Accessibility,    \newline Usability, \newline Legal Terms 
\end{noindlist}  
 &
  \begin{noindlist}
 		 \item Trust and social acceptance in vital towards the adaptation of the sensing as a service model. If sensor owners do not trust the sensing as a service, the entire model will fail. In order to win the trust, a long term change management process is required. It needs to be supported by increasing the awareness about inner-workings and benefits of the model. New privacy protection and security protocols \cite{P669} need to be introduced in order to make the model sustainable by winning the trust of all parties involved.
 		 
  		 \item Security and privacy is a must \cite{P670}. It needs to be implemented in number of levels. First, at the technology level, secondly, in government and business policy level and finally, through strict legal terms and conditions. Policies need to be set in order to keep the accessibility fairly open to the sensor data consumers while validating and monitoring all the parties involved in the model. Maximum usability at both ends (the sensor owner and sensor data consumer end) helps the model to be adopted by the wider community. Automated sensor configuration plays a significant role in usability because most of the sensor owner will be non-technical. 
 \end{noindlist}  

\\ \hline

\end{tabular}

\label{Tbl:Comparison of Semantic Technologies}
\vspace{-0.6cm}
\end{table*}



\section{Open Challenges}
\label{sec:Challenges}

The sensing as a service model can contribute significantly to address the challenges in the IoT and SC. There are many open challenges and issues that need to be tackled. We identify some of the major challenges in the sensing as a service model under three categories in Table \ref{Tbl:Comparison of Semantic Technologies}: technological, economical, and social, where some can be discussed under multiple categories. Each of these challenges shows research directions for future work in the sensing as a service domain.


\section{Conclusions}
\label{sec:Conclusions}

This paper provides a comprehensive overview of the sensing as a service model and its applicability towards Smart Cities in the Internet of Things paradigm. Our vision is backed up by a number of projects initiated around the globe, including FP7 ICT project OpenIoT \cite{P377}. We discussed the model from three different perspectives including technological, economical, and social. We examined how the sensing as a service can be a sustainable, scalable, and powerful model. The sensing as a service model allows utilizing  resources efficiently so limited resource can be used to accommodate large numbers of consumers. Further, it also creates a win-win situation for all the parties involved. We identified a number of major open challenges and issues which need to be addressed in order to realise the vision of sensing as a service. Finally, this model will create an unprecedented amount of opportunities to build innovative value added solutions that makes the decision making process efficient and effective in IoT paradigm.

\section*{Acknowledgement}

The authors acknowledge support from SSN TCP, CSIRO, Australia and ICT OpenIoT Project, which is co-funded by the European Commission under Seventh Framework program, contract number FP7-ICT-2011-7-287305-OpenIoT. The Author(s) acknowledge help and contributions from all partners of the OpenIoT project.


\bibliography{Bibliography}
\bibliographystyle{IEEEtran}


\end{document}